\documentclass[showpacs,prl,twocolumn,aps]{revtex4}
\usepackage{graphicx}
\usepackage{psfrag}
\usepackage{bm}
\begin{document}
\title{Dynamical role of anyonic excitation 
statistics in rapidly rotating Bose gases}
\author{Uwe R. Fischer}
\affiliation{Eberhard-Karls-Universit\"at T\"ubingen, 
Institut f\"ur Theoretische Physik \\  
Auf der Morgenstelle 14, D-72076 T\"ubingen, Germany}



\begin{abstract}                
We  show that for rotating harmonically trapped Bose gases 
in a fractional quantum Hall state, 
the anyonic excitation statistics in the rotating gas 
can effectively play a {\em dynamical} role.
For particular values of the two-dimensional coupling constant 
$g = -2\pi \hbar^2 (2k-1)/m$, where $k$ is a positive integer, 
the system becomes a noninteracting gas of anyons, 
with exactly obtainable solutions satisfying 
Bogomol'nyi self-dual order parameter equations.
Attractive Bose gases under rapid rotation thus can be stabilized 
in the thermodynamic limit due to the anyonic 
statistics of their quasiparticle excitations. 
\end{abstract}

\maketitle

{\em Introduction.}--Classical and quantum 
properties of atomic Bose gases when they are set
under rotation were intensely studied in the last couple of years  
both from the experimental 
\cite{VortexLatticeBEC,EngelsII,Stock} 
and the theoretical sides  
\cite{wilkin01,Paredes,Ho2001,Kasamatsu,EricJason,VortexLatticewGordon,CooperFeshbach,VortexQH}. 
Of particular interest is the behavior of 
rapidly rotating two-dimensional (2D) 
gases in a fractional quantum Hall state, 
corresponding to an electrically neutral, bosonic 
analog of ultrapure 2D electron gases in very strong magnetic fields, 
which exhibits strongly correlated physics \cite{wilkin01,Paredes}. 
It is well established 
that fractional quantum Hall states 
possess anyonic excitations above an incompressible ground state 
\cite{laughlin2},  
carrying certain fractions $\nu$ of the ``elementary'' charge  
\cite{wilczekfract}, where $\nu <1$ is the filling factor of 
the lowest Landau level. 
On the low-energy level of an effective field theory, these systems are
described by their coupling to a fictitious statistical gauge field 
$\cal A^\mu$.
This is known as a Chern-Simons effective field theory 
description of the fractional quantum Hall effect 
\cite{LopezFradkin91,EffectiveFTQH}.

In the thermodynamic limit, a system of untrapped 
bosons with attractive interaction is unstable against collapse and can 
stably exist only for a finite trapped number of particles 
\cite{nozieressaintjames,Bradley}.   
The purpose of the present contribution is to point out 
that a 2D harmonically trapped rotating Bose gas  
with attractive interactions can be stabilized 
by the statistical gauge field associated 
with the fractional charge of its quasiparticle excitations. 
The anyonic nature of the excitations
can therefore play a {\em dynamical} role, in that it can 
compensate the negative coupling constant associated to the 
interaction of atomic bosons. This dynamical role is due to the fact that 
the statistical ``magnetic'' field is proportional to the 
density of the system, resulting in particle-statistical flux composites.  
For a special value of the statistics parameter 
$\Theta = \nu/[2\pi (1-\nu)]$ or, correspondingly, for given $\Theta$, at 
a particular value of the coupling constant $g= -2\pi \hbar^2 (\nu^{-1}-1)/m$,  the interaction can effectively 
be eliminated entirely and make the system behave as an  
{\em interaction-free} gas of anyons. In physical terms, 
the stabilization of the attractively interacting gas in the 
fractional quantum Hall state against the 
collapse to a singular distribution may be understood 
by the existence of a 2D  analog of Fermi pressure, ``anyonic'' pressure. 

The order parameter distribution 
for the free anyon gas can be found analytically provided the Bogomol'nyi
self-duality identities are satisfied \cite{Bogomolnyi},  
which then leads to topological vortex solutions
for the Ginzburg-Landau order parameter  
\cite{jackiwpi,jackiwpiII,JackiwPiinB,Lin,Grossman,Dunne}. 
Rotating atomic Bose gases offer a unique opportunity to actually prepare and 
detect them rather directly and with great accuracy, and to 
study their {\em breather soliton} dynamics \cite{jackiwpiII,JackiwPiinB}.
Starting from the statistical mechanics of particles in the plane, 
interacting by contact potentials,  
other investigations on obtaining a free anyon gas 
can be found in \cite{bhaduri}; however, only 
repulsive interactions were treated, without externally
imposed magnetic or rotation fields.

 {\em Abelian statistical gauge field.}--We assume that we have a 2D gas  
in the fractional quantum Hall regime 
which admits a Ginzburg-Landau type 
description in terms of a Chern-Simons theory \cite{EffectiveFTQH}.  
A {statistical} (2+1)D gauge three-potential ${\cal A}^\mu$ 
may be implemented by showing the 
physical equivalence of the two Hamiltonian theories with ${\cal A}^{\mu}=0$ 
and ${\cal A}^{\mu}\neq 0$ on the quantum level \cite{LopezFradkin91}.
The primary effect of the statistical 
vector potential is that it allows for an additional Chern-Simons 
term in the low-energy action ($\hbar \equiv 1$):  
\begin{eqnarray}
S_\Theta & = & \int d^2 {\bm x} dt   \left( i\Psi^{*} (\partial_t +i{\cal A}_0)\Psi 
- \frac1{2m} |{\bm D} \Psi|^2\right.
\nonumber\\ 
 & & \hspace*{1.5em}\left.  - V({\bm x}) |\Psi|^2 
-{\cal U} (|\Psi|^2) 
+\frac\Theta 4  
\epsilon^{\alpha\beta\gamma} {\cal A}_{\alpha} {\cal F}_{\beta\gamma} 
\right). \label{action} 
\end{eqnarray}
Here, $\bm D$ is the  gauge covariant spatial derivative
defined in Eq. (\ref{D}), the Abelian field strength of 
the statistical gauge field reads 
${\cal F}_{\alpha\beta} =\partial_\alpha {\cal A}_\beta 
- \partial_\beta {\cal A}_\alpha$, 
and ${\cal U} (|\Psi|^2)$ is the 
interaction energy density. We use relativistic notation,  
i.e.,  $\alpha,\beta,\gamma,\mu = 0,1,2$; the spacetime 
metric is diag\,$(1,-1,-1)$ and 
repeated indices are summed over.
All higher derivative terms in the statistical 
gauge potentials 
(e.g., Maxwell terms $\propto {\cal F}_{\alpha\beta} {\cal F}^{\alpha\beta}$)
are left out in this low-energy, low-momentum expression \cite{Maxwell}. 

Classically, the momenta of the particles  
making up the system are given by the decomposition 
$ 
{\bm p} = m {\bm v} + \kappa{\bm A} + {\cal {\bm A}}, \label{p}
$ (the spatial part of ${\cal A}^\mu$ is denoted  ${\cal A}$),  
so that the gauge covariant derivative in the quantum domain is given by
\begin{equation}
{\bm D} = \nabla - i\kappa {\bm A} - i{\cal A}. \label{D}
\end{equation} 
Here, ${\bm A}$ is the external U(1) 
vector potential of rotation, yielding the Coriolis force, 
with $\nabla\times {\bm A} = 2{\Omega}$
twice the applied rotation field $\Omega$ 
perpendicular to the 2D plane,  
and $\kappa=m$ is the coupling constant for a rotating Bose gas. 
The total scalar potential 
\begin{equation}
V = V_{\rm trap} - \frac12 m\Omega^2 r^2  
\end{equation} 
consists of (harmonic) trapping potential and centrifugal potential. 
Due to the Chern-Simons term, the field strength of the statistical gauge field
is given by the current as follows: 
\begin{equation}
\frac\Theta 2 \epsilon^{\mu\alpha\beta} {\cal F}_{\alpha\beta} = J^{\mu}.
\label{Gauss} 
\end{equation} 
The homogeneous Maxwell equation for the statistical field 
strength, 
$\partial_\mu {\cal F}_{\alpha\beta}+\partial_\beta {\cal F}_{\mu\alpha}
+\partial_\alpha{\cal F}_{\beta\mu}=0$, 
is then automatically equivalent to 
current conservation, $\partial_\mu J^\mu=0$. 

The statistical gauge field strength is according to 
(\ref{Gauss}) {\em dual} to the current. 
This means, in particular, that the statistical magnetic field is
proportional to the density:  
\begin{equation}
-\Theta {\cal B} = |\Psi|^2 = \rho . \label{calBdensity}
\end{equation} 
That is, the order parameter modulus is inextricably linked to 
the statistical flux, and particle-flux composites are formed.  
Taking $\Theta > 0$, by fixing $\Omega >1$, the statistical magnetic field 
cancels part of the applied ``magnetic'', i.e., rotation field in the 
canonical momentum $\bm p$, 
and the vector potential effectively acting on the particles is reduced.
Defining the Landau level filling factor of the original bosonic 
particles to be $\nu =\pi \rho_0 /m \Omega $ \cite{wilkin01,VortexQH}, with 
$\rho_0 = |\Psi_0|^2$ 
a homogeneous background density, we have $\Theta = \nu /[2\pi (1-\nu)]$.
In particular,
for the $\nu = 1/2$ anyons discussed in \cite{Paredes}, $\Theta = 1/2\pi$.
The effective magnetic flux of particle-statistical flux composites   
$\tilde \Phi  = \oint d{\bm x}\cdot [{\bm A} +{\cal A}/m]=\nu\Phi_0$, then 
is reduced compared to the bare  rotational flux quantum 
$\Phi_0 = \oint d{\bm x}\cdot {\bm A} = 2\pi /m$ obtained for 
vanishing $\cal A$ \cite{arovaswilczek}.

The Ginzburg-Landau energy functional of the gas in the rotating 
frame, corresponding to (\ref{action}), is composed as usual of 
kinetic, scalar potential and interaction energy,  
\begin{equation}
H = \int d^2 {\bm x}  \left( \frac1{2m}
|{\bm D} \Psi|^2 + V({\bm x}) |\Psi|^2 + {\cal U} (|\Psi|^2)
\right) .
\label{H}
\end{equation}
In the rapid rotation limit $\Omega \simeq \omega_\perp$, 
$V_{\rm trap} = \frac12 m\omega_\perp^2 r^2$ is (nearly) cancelled
by the centrifugal potential, $V\simeq 0$, 
where $\omega_\perp$ is the trapping
frequency perpendicular to the axis of rotation. 

Using the Bogomol'nyi decomposition \cite{Bogomolnyi} 
for the kinetic energy term in the integral (\ref{H}),  
\begin{equation}
|{\bm D} \Psi|^2 = |(D_1\pm i D_2) \Psi|^2
\pm \nabla\times {\bm J} \pm {\cal B} \rho  \pm 2m {\Omega} \rho,  
\label{D1D2}
\end{equation}
and relation (\ref{calBdensity}), we can 
rewrite 
the Hamiltonian 
in two different ways corresponding to the $\pm $ 
sign in (\ref{D1D2}), to read
\begin{eqnarray} 
H & = & 
\int d^2 {\bm x}  \left[ \frac1{2m}
|(D_1 \pm i D_2) \Psi|^2 
+ V ({\bm x}) |\Psi|^2 
\right.\nonumber\\& & \left. \hspace*{3em}
+  {\cal U} (|\Psi|^2) 
\mp \frac12 \left(\frac1{m \Theta}|\Psi|^2   - 2  {\Omega} \right) 
|\Psi|^2
\right]\!. 
\label{H2}
\end{eqnarray}
We have neglected the term involving the curl of the 
matter current $\pm\nabla\times {\bm J}$, as it amounts, after integration,  
to a surface contribution vanishing for sufficiently well-behaved 
current fields and/or well outside the boundary of the rotating gas.  
We infer from the last term in the second line of (\ref{H2})  
that due to the proportionality
of density and statistical magnetic field expressed by 
Eq.\,(\ref{calBdensity}), and employing the identity (\ref{D1D2}), 
part of the kinetic energy $|{\bm D} \Psi|^2/2m$ can effectively 
act as interaction energy,

The potential in the Ginzburg-Landau energy functional may generally
be expanded 
\begin{equation}
{\cal U} (|\Psi|^2) 
= \frac{g}2 |\Psi|^4 + \frac{\gamma}6 |\Psi|^6 + \cdots,
\end{equation}
where $g$ is the 
value of the two-body coupling 
constant in the fractional quantum Hall state; for sixth order 
stability of the system, 
$\gamma > 0$ is required. Taking into account the coefficient 
of the quartic term in (\ref{H2}) resulting from the above expansion and 
Eqs.\,(\ref{Gauss}) and (\ref{D1D2}),  
the effective interaction coupling of the system is now defined as  
\begin{equation}
g_{\rm eff} = g \mp \frac1 {m\Theta}.  \label{geff}
\end{equation} 
If we choose the relation of coupling constant and 
statistics parameter to be 
$g = \pm 1/ m{\Theta}$, 
we see 
that we have effectively eliminated the quartic interaction coupling term. 
By ``interaction-free'' 
we mean that we have cancelled the interaction $g$ by a proper choice of 
the statistical magnetic field $\cal B$, while the covariant derivatives 
$D_1$ and $D_2$ do of course still implicitly 
contain the statistical gauge fields ${\cal A}_1$ and ${\cal A}_2$, 
and therefore, due to (\ref{Gauss}),  the density distribution. 

We conclude from relation (\ref{geff}) that 
a {large} original coupling $g$ can be converted to a $g_{\rm eff}$ 
approaching zero. 
Due to the statistical interaction, 
a negative $g$ system, unstable towards collapse 
in the thermodynamic limit \cite{nozieressaintjames}, 
may be stabilized by the statistical interaction, 
in the sense that there are nonsingular distributions  
$\Psi ({\bm x},t)$ (see below), 
with finite energy. 
The value $g = g_c \equiv -1/m\Theta$ 
defines a critical negative coupling strength, below which the gas
cannot be stabilized 
for a given $\Theta$.

In the limit that the coefficient 
of $|\Psi|^4$ vanishes, $g_{\rm eff}=0$ in (\ref{geff}), 
the problem of determining 
the ground state of zero energy becomes exactly solvable (neglecting  
the small $|\Psi|^6$ term in the Ginzburg-Landau expansion).  
The energy per particle 
then assumes its lower limit $H_0/N  = \pm \Omega $, 
provided the {\em self-duality} (Bogomol'nyi) constraints 
\cite{Bogomolnyi} 
\begin{equation}
(D_1 \pm i D_2) \Psi =0  \label{selfduality}
\end{equation}
are satisfied. 
To protect the self-dual states of effectively zero coupling,  
there is an energy barrier to neighboring states in $\Psi$ space, 
which do not have the property that $g_{\rm eff}=0$.
The magnitude of this energy barrier depends 
on the (positive) contribution of the 
$|\Psi|^6$ term in the Ginzburg-Landau type expansion 
of ${\cal U} (|\Psi|^2)$.

Jackiw and Pi \cite{jackiwpi} 
have shown that solving the self-duality equations (\ref{selfduality})
above, in the case of zero ``external'' field, i.e., in 
the problem with just the statistical gauge field present
and trapping and rotation turned off, $V= 0$ 
and $\Omega =0$, 
is equivalent to 
solving the Liouville equation  
\begin{equation}
\Delta \ln \rho = \pm  2\rho/\Theta \qquad (V({\bm x})=\Omega=0), \label{Liouville}
\end{equation}
whose complete set of solutions is known \cite{Wirtinger}. 
To obtain regular and non-negative
solutions of the Liouville equation, the sign  of the right-hand side
must be chosen opposite to that of $\Theta$. 
Thus the lower, minus sign is appropriate for $\Theta >0$.  
The general solution of the Liouville 
Eq.\,(\ref{Liouville}) 
in the plane of complex $z= x + iy$ reads 
$
{\rho(z})= {4\Theta |f'(z)|^2}/{[1+ |f(z)|^2]^2} ,
$ 
where $f(z)$ is an arbitrary holomorphic function. 
Radially symmetric vortex solutions, for 
which $\Psi =\sqrt{\rho} \exp[in\phi]$, 
take the form, choosing $f(z)= (z_0/z)^n$,  
\begin{equation}
\rho(r) = \frac{4\Theta n^2}{r^2} \left[\left(\frac{r_0}{r}\right)^n
+\left(\frac r{r_0}\right)^n \right]^{-2}, \label{rhonr}
\end{equation}
where $n$ is an integer, and $r_0$ an arbitrary length scale reflecting 
the scale (dilation) invariance of (\ref{Liouville}); 
for $r\rightarrow 0$, $\rho (r) \propto r^{2(n-1)}$, and 
for $r\rightarrow \infty$, $\rho (r) \propto r^{-2(n+1)}$.

The solutions of (\ref{selfduality}) 
with a constant external magnetic/rotation 
field (corresponding to our being in the rotating frame), 
and a linearly increasing electric field in the plane 
(corresponding to our linear trapping and centrifugal forces) 
can be obtained from the vortex solitons of Jackiw and Pi: 
Adding these additional external fields the problem (still) is quadratic 
\cite{jackiwpiII,JackiwPiinB,Lin}. 
The scalar potential in the Hamiltonian (\ref{H2}), i.e.
the effective harmonic trapping field, is very small 
close to criticality, 
$V=\frac12 m\omega_\perp^2 r^2 -\frac12 m \Omega^2 r^2 \simeq 0$, and can 
be neglected. The classical problem of finding the solution of 
(\ref{selfduality}) in an external rotation field  then corresponds 
to the problem of finding the (semiclassical) Landau levels 
of anyons \cite{jackiwpiII}.

To generate time-dependent vortex soliton solutions, 
one constructs a coordinate transformation, whose inverse 
effectively removes the external field and thus leads us back to (\ref{rhonr})
\cite{jackiwpiII,JackiwPiinB,Lin}. The most important feature, apart from 
the cyclotron motion executed by the soliton, 
which results from this transformation, is that the 
size of the soliton {\em breathes} if the background $\Psi_0 \neq 0$: 
The scale factor in (\ref{rhonr}) becomes time dependent 
according to $r_0 \rightarrow  r_0 \cos [\Omega t ]$. Hence the 
soliton size oscillates with the applied rotation  frequency. 
Furthermore, the  energy of the soliton diverges for $n=1$ and 
finite $r_0$, and thus according to (\ref{rhonr}) only $n\ge 2$ configurations
with $\rho= 0$ at the vortex line can be generated; note that
for $n=1$ the density (\ref{rhonr}) at the origin has a finite value.  
The statistical vector potential ${\cal A}$
decreases at large distances from the center of the vortex line like $1/r$, 
as required for a topological vortex of a given quantized circulation. 
It should also be stressed that the feature that there exists a breather 
soliton solution in the presence of the external rotation field, persists 
when $g\neq g_c$ and there is no self-duality fulfilled according to 
(\ref{selfduality}); however, in that case simple analytical solutions like
the one displayed in (\ref{rhonr}) cannot be obtained, because the 
Ginzburg-Landau equations then remain essentially nonlinear. 

Rescaling the coordinate vector via ${\bm x}= 
\tilde {\bm x}\sqrt{\Theta/2\rho_0}$, 
the typical length scale of inhomogeneous solutions 
of the equations
(\ref{selfduality}) 
is set by $\xi_0 = \sqrt{\Theta/2\rho_0} = \sqrt{1/2m\rho_0 |g|}$, the
analog of the coherence length in the repulsive case. 
The topological, i.e., quantized circulation   
vortex solutions in a rapidly rotating gas with $g<0$, 
resulting from the boosted solutions (\ref{rhonr}), 
then are cousins of their counterparts 
in the repulsive-interaction superfluid.  
In the latter case, 
the core size of the 
vortices depends on the applied rotation rate 
\cite{VortexLatticewGordon} (for an 
experimental verification see the second reference 
of \cite{EngelsII}). Here, by contrast, 
the scale $r_0$ in Eq.\,(\ref{rhonr}) is essentially a free parameter, because the
total energy of the soliton is minimized for $r_0= 0$ \cite{JackiwPiinB}.
Therefore, an external force, created by a blue-detuned laser for example, 
has to be applied to the gas to generate a soliton with a 
finite value of $r_0$.
It should be noted 
that $\Omega \neq 0$ is strictly necessary 
to obtain any solution of nonzero $\Psi_0$, i.e.,  
the symmetry breaking expressed by 
$\Psi_0 \neq 0$ is {\em induced} by the
rotation of the gas: The solutions of (\ref{Liouville}), e.g., the 
vortex solution in Eq.\,(\ref{rhonr}),  are all 
asymptotically vanishing, $\Psi_0=0$ if $\Omega =0$.

{\em Non-Abelian case.}--The previous considerations can be generalized to the case
of a noncommuting statistical gauge field (cf., e.g., \cite{Grossman,Dunne}), 
which is of potential relevance in rapidly rotating spinor gases.
An outline of such a non-Abelian generalization reads as follows. 
The non-Abelian field strength is 
\begin{equation}
{\cal F}_{\alpha\beta} = \partial_\alpha {\cal A}_\beta 
-\partial_\beta{\cal A}_\alpha + [{\cal A}_\alpha,{\cal A}_\beta]
\end{equation} 
where the noncommuting 
gauge field reads ${\cal A}_\mu = {\cal A}^a_\mu T^a$,  
using the Lie algebra of the anti-Hermitian generators $T^a$ of  
the non-Abelian group, $[T^a,T^b]= f_{abc} T^c$ (summation over $a,b,c$ 
is implied; $f_{abc}$ are the structure constants of the Lie algebra). 
The Chern-Simons-Gauss law (\ref{calBdensity}) 
reads in its non-Abelian form  
\begin{eqnarray}
{\cal B}
= \frac{i}{\Theta} T^a ({\bm \psi}^\dagger T^a {\bm \psi}), \label{nonabelB} 
\end{eqnarray}
where $\bm \psi$ is the order parameter field, transforming 
according to some given representation of the gauge group. 

The scalar and vector 
interactions in a spinor 
condensate may be parametrized to read 
\cite{Ohmi,EisenbergLieb}  
\begin{equation}
V_{\rm int} = \frac{c_0}2 \psi^\dagger_i \psi^\dagger_j \psi_j \psi_i 
- \frac{c_2}2 \psi^\dagger_i \psi^\dagger_k T^a_{ij} T^a_{kl} \psi_l \psi_j , 
\label{SpinorV}
\end{equation}
where the summation indices $i,j,k,l$ 
cover the matrix index of the particular 
representation of $\bm\psi$ chosen. 
We infer, using (\ref{nonabelB}), that the effective coefficient of the 
quartic spin-spin interaction term may in analogy to (\ref{geff}) be 
defined, restricting ourselves to the lower sign,   
\begin{equation}
(c_2)_{\rm eff}= c_2 + \frac{1}{m\Theta}. 
\end{equation}
The statistical interaction therefore is able to change, 
assuming $c_0<0$, the spin-spin interaction from $c_2<0$ (ferromagnetic)
to $c_2>0$ (``polar'' \cite{Ohmi}).
However, for a two-parameter interaction Hamiltonian like (\ref{SpinorV}),  
with $c_0 \neq 0$, 
there is no self-dual Bogomol'nyi 
point in parameter space. Only at 
$c_0=0$, such a point exists; then, the system,   
in general,  
can support vortices obeying non-Abelian 
fractional statistics. The matter density components associated 
to these fractional vortices are solutions of the Toda equation, 
which generalizes the Liouville
Eq.\,(\ref{Liouville}) \cite{Dunne,Grossman}.

{\em Conclusion.}--We have established the fact 
that a rapidly rotating Bose gas with attractive self-interaction can be made
manifestly stable in a given fractional quantum Hall state, such that 
for certain values of the negative 
coupling constant the Ginzburg-Landau matter wave field
exhibits self-duality.  
The resulting free anyon gas 
interpolates between the extreme cases of 
bosons (zero statistical pressure at $T=0$) and fermions
(maximal statistical pressure at $T=0$), and 
the original bosonic gas is protected against collapse because of this 
effective ``anyonic'' pressure.  
The preparation and 
experimental verification of this particular fractional quantum Hall 
state should, in principle,  
be possible starting from a rotating Bose gas with 
small positive interaction coupling and high angular momentum 
$L= N(N-1)/2\nu = {\cal O} (N^2)$. To avoid the problem of 
stabilizing a bulk gas at these very large angular momenta, one can conceive
of putting the system on an optical lattice \cite{Popp}. 
Starting from the small positive interaction coupling Bose gas, 
one switches 
over a close-lying Feshbach resonance to  
the negative coupling strength side of the resonance.
At the end point of the sweep a coupling is chosen which  
fulfills $g= -2\pi (2k-1) /m$, 
corresponding to a 
fractional quantum Hall state at filling $\nu=1/2k$.

I gratefully acknowledge discussions with Petr O. Fedichev and Nils Schopohl.


\begin{thebibliography}{499}
\bibitem{VortexLatticeBEC} J. R. Abo-Shaeer, C. Raman, J. M. Vogels, and 
W. Ketterle, 
Science {\bf 292}, 476 (2001).
\bibitem{EngelsII} P. Engels {\it et al.}, 
Phys. Rev. Lett. {\bf 90}, 170405 (2003); 
V. Schweikhard {\it et al.}, 
Phys. Rev. Lett.  {\bf 92}, 040404 (2004). 
\bibitem{Stock} V. Bretin, S. Stock, Y. Seurin, and 
J. Dalibard, 
Phys. Rev. Lett. {\bf 92}, 050403 (2004); 
S. Stock {\it et al.},
Europhys. Lett. {\bf 65}, 594 (2004).  
\bibitem{wilkin01} N.\,R. Cooper, N.\,K. Wilkin, and J.\,M.\,F. Gunn,
Phys. Rev. Lett. {\bf 87}, 120405 (2001).  
\bibitem{Paredes} B. Paredes, P. Fedichev, J.\,I. Cirac, and P. Zoller,
Phys. Rev. Lett. {\bf 87}, 010402 (2001); 
B. Paredes, P. Zoller, and J.\,I. Cirac, 
Phys. Rev. A {\bf 66}, 033609 (2002). 
\bibitem{Ho2001} T.-L. Ho, Phys. Rev. Lett. {\bf 87}, 060403 (2001).
\bibitem{Kasamatsu} K. Kasamatsu, M. Tsubota, and M. Ueda, 
Phys. Rev. A {\bf 66}, 053606 (2002); Phys. Rev. Lett. {\bf 91}, 150406 (2003).
\bibitem{EricJason} E.\,J. Mueller and T.-L. Ho, 
Phys. Rev. Lett. {\bf 88}, 180403 (2002);  
Phys. Rev. A {\bf 67}, 063602 (2003).
\bibitem{VortexLatticewGordon} U.\,R. Fischer and G. Baym,  
Phys. Rev. Lett. {\bf 90}, 140402 (2003); 
G. Baym and C.\,J. Pethick,
Phys. Rev. A {\bf 69}, 043619 (2004); 
G. Watanabe, G. Baym, and C.\,J. Pethick, cond-mat/0403470  
[Phys. Rev. Lett. (to be published)].
\bibitem{CooperFeshbach} N.\,R. Cooper,
Phys. Rev. Lett. {\bf 92}, 220405 (2004).
\bibitem{VortexQH} U.\,R. Fischer, P.\,O. Fedichev, and A. Recati,
J. Phys. B 
{\bf 37}, S301 (2004).
\bibitem{laughlin2} R.\,B. Laughlin, 
Phys. Rev. Lett. {\bf 50}, 1395 
(1983).
\bibitem{wilczekfract} F. Wilczek,
Phys. Rev. Lett. {\bf 49}, 957 (1982).
\bibitem{LopezFradkin91} A. Lopez and  E. Fradkin, 
Phys. Rev. B {\bf 44}, 5246 
(1991). 
\bibitem{EffectiveFTQH} S.-C. Zhang, T. H. Hansson, and S. Kivelson, 
Phys. Rev. Lett. {\bf 62}, 82 (1989);  
N. Read, 
Phys. Rev. Lett. {\bf 62}, 86 (1989); 
S.-C. Zhang, Int. J. Mod. Phys. B {\bf 6}, 25 (1992).
\bibitem{nozieressaintjames} P. Nozi\`eres and D. Saint James,
J. Physique {\bf 43}, 1133 (1982).
\bibitem{Bradley} C.\,C. Bradley, C.\,A. Sackett, and R.\,G. Hulet,
Phys. Rev. Lett. {\bf 78}, 985 (1997).
\bibitem{Bogomolnyi} E.\,B. Bogomol'nyi, Sov. J. Nucl. Phys. {\bf 24}, 449 
(1976) [Yad. Fiz. {\bf 24}, 861 (1976)].  
\bibitem{jackiwpi} R. Jackiw and S.-Y. Pi, 
Phys. Rev. Lett. {\bf 64}, 2969 (1990). 
 \bibitem{jackiwpiII} R. Jackiw and S.-Y. Pi, 
Phys. Rev. Lett. {\bf 67}, 415 (1991).
\bibitem{JackiwPiinB}  Z. F. Ezawa, M. Hotta, and A. Iwazaki, 
Phys. Rev. Lett. {\bf 67}, 411 (1991);  
Phys. Rev. D {\bf 44}, 452 (1991).
\bibitem{Lin} R. Jackiw and S.-Y. Pi, Phys. Rev. D {\bf 44}, 2524 (1991).
\bibitem{Grossman} B. Grossman, Phys. Rev. Lett. {\bf 65}, 3230 (1990). 
\bibitem{Dunne} G.\,V. Dunne, R. Jackiw, S.-Y. Pi, and C.\,A. Trugenberger, 
Phys. Rev. D {\bf 43}, 1332 (1991).
\bibitem{bhaduri} R.\,K. Bhaduri, M.\,V.\,N. Murthy, and M.\,K. Srivastava, 
Phys. Rev. Lett. {\bf 76}, 165 (1996); 
T.\,H. Hansson, J.\,M. Leinaas, and S. Viefers, 
Phys. Rev. Lett. {\bf 86}, 2930 (2001). 
\bibitem{Maxwell} If Maxwell terms 
are included, the Chern-Simons term, whose inclusion is consistent 
in 2D only, results in a planar ``electrodynamics'' where the ``photons'' 
have mass, yet for which U(1) gauge invariance is maintained. 
\bibitem{arovaswilczek} D. Arovas, J.\,R. Schrieffer, and F. Wilczek,
Phys. Rev. Lett. {\bf 53}, 722 (1984).
\bibitem{Wirtinger} W. Wirtinger, Math. Ann. {\bf 97}, 357 (1927). 
\bibitem{Ohmi} T. Ohmi and K. Machida,
J. Phys. Soc. Jpn. {\bf 67}, 1822 (1998); 
T.-L. Ho, 
Phys. Rev. Lett. {\bf 81}, 742 (1998).
\bibitem{EisenbergLieb} Note 
it has been shown [E. Eisenberg and E.\,H. Lieb,
Phys. Rev. Lett. {\bf 89}, 220403 (2002)], 
that the interaction $V_{\rm int}$, 
displayed in Eq. (\ref{SpinorV}), 
contains the $c_2$ spin-spin term
only if the 
{\em microscopic} Hamiltonian has explicitly spin-dependent interactions.
\bibitem{Popp} M. Popp, B. Paredes, and J.\,I. Cirac, 
cond-mat/0405195 [Phys. Rev. A (to be published)]. 
\end{thebibliography}
\end{document}